\documentstyle[12pt]{article}
\def\be{\begin{equation}}
\def\ee{\end{equation}}
\def\bea{\begin{eqnarray}}
\def\eea{\end{eqnarray}}

\begin{document}
\begin{center}
\Large
{\bf Classical gravitational spin-spin interaction}
\end{center}
\vspace{.3in}
\normalsize
\begin{center}
{\sc W.B.Bonnor}\\
\end{center}
\normalsize
\vspace{.6cm}
\begin{center}
{\em School of Mathematical Sciences, \\
Queen Mary, University of London,
\\Mile End Road, \\ London. E1 4NS. UK \\}
\vspace{0.3in}
\end{center}
\begin{abstract}
I obtain an exact, axially symmetric, stationary solution of Einstein's equations for two massless
spinning particles.  The term representing the spin-spin interaction agrees with
recently published approximate work.  The spin-spin force appears to be proportional
to the inverse fourth power of the coordinate distance between the particles.
\end{abstract}

\section{Introduction}
Papers on exact stationary solutions of Einstein's equations for
pairs of massive spinning particles have been published by several authors
[1-8].
These, though ingenious and important, particularly in the study of the
equilibrium of black holes, contain no general formulae for the interactions
between the particles.
Recently I have studied this problem [10]
by a method of approximation and obtained
some general formulae which include interaction terms up to the second approximation in
the masses and angular momenta of the particles.

There exists a class of exact solutions which offers important information on this
subject in a surveyable form.  This is the  Papapetrou class [9], which
is stationary and axially symmetric, and depends on one harmonic function.
Using it one can construct a solution representing two particles bearing
angular momentum but no mass.  Because of the masslessness the solution is incomplete,
of course, but nevertheless it gives an exact account of spin-spin gravitational
interaction.  This provides a check on the interaction obtained
in [10].  The agreement turns out to be satisfactory, which
boosts one's confidence in the approximation method of [10].

The plan of the paper is as follows. In section 2 I give the metric and
the field equations, and in section 3 I describe Papapetrou's class of solutions
and specialise it to the case of two spinning particles.
The physical interpretation is discussed in section 4 where I give an
estimate of the spin-spin force, and there is a concluding section 5.

\section{Metric and field equations}
We start from the metric for a stationary axially symmetric spacetime:
\begin{equation}
ds^{2}=-f^{-1}e^{\nu}(dz^{2}+dr^{2})-l d\theta^{2}-2n d\theta dt+fdt^{2},
\end{equation}
where $f,\nu,l,n$ are functions of $r$ and $z$.  We use the vacuum field
equations, sources being represented by singularities, and
number the coordinates
\[x^{1}=z, x^{2}=r, x^{3}=\theta, x^{4}=t,   \]
where
\[ \infty>z>-\infty,\:r>0,\:2\pi\geq\theta\geq0,\:\infty>t>-\infty, \]
$\theta=0$ and $\theta=2\pi$ being identified.
Proceeding as in [10], we find that the field equations
\[R_{ik}=0 \]
can be put
in the form
\begin{eqnarray}
R_{11}+R_{22}=\nu_{11}+\nu_{22}-f^{-1}\nabla^{2}f+\frac{3}{2}f^{-2}(f_{1}^{2}+f_{2}^{2})\nonumber\\-\frac{1}{2}r^{-2}f^{2}(w_{1}^{2}+w_{2}^{2})&=&0,\\
R_{11}-R_{22}=r^{-1}\nu_{2}+\frac{1}{2}f^{-2}(f_{1}^{2}-f_{2}^{2})+\frac{1}{2}r^{-2}f^{2}(w_{2}^{2}-w_{1}^{2})&=&0,\\
2R_{12}=-r^{-1}\nu_{1}+f^{-2}f_{1}f_{2}-r^{-2}f^{2}w_{1}w_{2}&=&0,\\
R_{4}^{4}-R_{3}^{3}-2wR_{4}^{3}=e^{-\nu}[-\nabla^{2}f+f^{-1}(f_{1}^{2}+f_{2}^{2})-r^{-2}f^{3}(w_{1}^{2}+w_{2}^{2})]&=&0,\\
2R_{4}^{3}=-r^{-2}f^{2}e^{-\nu}[f\nabla^{*2}w+2(f_{1}w_{1}+f_{2}w_{2})]&=&0,
\end{eqnarray}
where suffices 1 and 2 on the right denote $\partial /\partial z$ and $\partial/ \partial r$, respectively, and
\begin{eqnarray*}
w&=&nf^{-1},\\
\nabla^{2}X&=&X_{11}+X_{22}+r^{-1}X_{2},\\
\nabla^{*2}X&=&X_{11}+X_{22}-r^{-1}X_{2},
\end{eqnarray*}
and $l$ is given by
\begin{equation}
lf+n^{2}=r^{2}.
\end {equation}

\section{The exact solution}
Papapetrou's class of solutions of (2)-(7) is
\footnote{One may take $f^{-1}=\alpha \cosh\xi_{1}+\beta \sinh\xi_{1},$
where $\alpha$ and $\beta$ are constants.  I put $\beta=0$ to exclude
unphysical mass dipoles, and $\alpha=1$ to ensure $f=1$ at infinity.}
\begin{eqnarray}
f^{-1}&=&\cosh \xi_{1},\\
n&=&rf\xi_{2},\\
\nu_{1}&=&r\xi_{11}\xi_{12},\\
\nu_{2}&=&\frac{1}{2}r[(\xi_{12})^{2}-(\xi_{11})^{2}],\\
\nabla^{2}\xi&=&0,\\
l&=&f^{-1}(r^{2}-n^{2}),
\end{eqnarray}
where $\xi$ is a function of $z$ and $r$, and
a subscript $1$ or $2$ means differentiation as described above.

A solution for two particles on the $z$-axis at $z=\pm b$, with spins parallel or antiparallel,
arises if we choose
\begin{equation}
\xi=-2\left(\frac{h_{1}}{R_{1}}+\frac{h_{2}}{R_{2}}\right),
\end{equation}
where $h_{1}$ and $h_{2}$ are constants describing the angular momenta of the particles,
and $R_{1}=\mid[(z-b)^{2}+r^{2}]^{1/2}\mid, R_{2}=\mid[(z+b)^{2}+r^{2}]^{1/2}\mid$.
This gives
\begin{eqnarray}
f^{-1}&=&\cosh2\left(\frac{h_{1}(z-b)}{R_{1}^{3}}+\frac{h_{2}(z+b)}{R_{2}^{3}}\right),\\
n&=&2fr^{2}\left(\frac{h_{1}}{R_{1}^{3}}+\frac{h_{2}}{R_{2}^{3}}\right),\\
\nu&=&\sum_{i=1}^{2}\frac{h_{i}^{2}r^{2}(9r^{2}-8R_{i}^{2})}{2R_{i}^{8}}
-\frac{h_{1}h_{2}[3(r^{2}+z^{2}-b^{2})^{3}+2b^{2}r^{2}(9r^{2}+9z^{2}-b^{2})]}{2b^{4}R_{1}^{3}R_{2}^{3}}\nonumber \\ +C,
\end{eqnarray}
where $C$ is an arbitrary constant.  $l$ can be obtained from (7),(15) and (16).

\section{Physical interpretation}
We see that $n$ represents the spin term for two particles of angular momenta
$h_{1}, h_{2}$ at points $z=\pm b$ on the $z$-axis.  However, expanding $f$
in inverse powers of $R_{1}$ and $R_{2}$ we have
\[f=1-2\left(\frac{h_{1}(z-b)}{R_{1}^{3}}+\frac{h_{2}(z+b)}{R_{2}^{3}}\right)^{2}+O(R^{-8}), \]
where $R^{2}=z^{2}+r^{2}$.  This contains no term of order $R^{-1}$, i.e.
no term representing mass.  Thus we are modelling massless spinning particles;
nevertheless, our solution will describe the spin-spin interaction.  I shall
comment further on this in the Conclusion.

The spin-spin interaction appears in the expression for $\nu$.
For Euclidean geometry on the axis we need
\begin{equation}
\lim_{r\rightarrow 0}\nu=0,
\end{equation}
and so from (17),
\begin{equation}
C- \frac{3h_{1}h_{2}(z^{2}-b^{2})^{3}}{2b^{4}R_{1}^{3}R_{2}^{3}}=0.
\end{equation}
The arbitrary constant $C$ can be chosen to satisfy this either for $\mid z \mid<b$
or for $\mid z \mid> b$, but not both.   Let us choose
\begin{equation}
C=\frac{3h_{1}h_{2}}{2b^{4}},
\end{equation}
so that (19) is satisfied for $\mid z \mid>b$; then there is a conical singularity
for $\mid z \mid<b$, i.e. between the particles.  I interpret this in the usual
way as a massless rod holding the particles in position, countering
the spin-spin interaction.

The value (20) of $C$ is the same as the spin-spin part of the corresponding
constant in [10,eqn (28)], which adds  to the credibility of the approximation
method used there.

To quantify the spin-spin force I proceed as follows.
Consider the metric near $r=0$ for $\mid z \mid<b$.  Rescaling the coordinates
by introduction of $\rho$ and $Z$ by
\[\rho=re^{C}, Z=ze^{C}, \]
one obtains, neglecting terms of order $\rho^{2}$ except in $g_{\theta\theta}$,
\begin{equation}
 ds^{2}=-f_{0}^{-1}(dZ^{2}+d\rho^{2}+\rho^{2}e^{-2C} d\theta^{2})+f_{0} dt^{2}, .
\end{equation}
where $f_{0}=\lim_{r\rightarrow 0}f, \mid z \mid <b$, so $f_{0}$ depends
on $z$.

If $f_{0}$ were a constant (21) would be the metric for a (finite) cosmic
string.  Provided $h_{1}$ and $h_{2}$ are small, and the point on the axis
under consideration is not too near the particles, $f_{0}$ is approximately
equal to unity, so let us assume that the conical singularity in (21) does
represent a cosmic string.  Then its linear density is $\lambda=\frac{1}{4}(1-e^{-C})$
[11].  The equation of
state of the cosmic string is $p+\lambda=0$, and it exerts no gravitational
field, but it does exert a force \footnote {In units of customary dimensions
the force is $(-3Gh_{1}h_{2})/(8c^{2}b^{4})$.} on the particles at its ends:
\begin{eqnarray}
p&=&-\frac{1}{4}(1-e^{-C}) \sim -\frac{1}{4}C\\
 &=&-\frac{3h_{1}h_{2}}{8b^{4}},
\end{eqnarray}
which is a measure of the spin-spin force, repulsive if the spins are
parallel.

This formula is no more than suggestive.  Moreover, it cannot be given a
rigorous meaning in terms of proper distance
as the segment $\mid z\mid<b$ of the axis is singular.  Note, however,
that in the approximate calculations of [10], which take account of the ordinary
gravitational attraction of the particles, the constant $\stackrel{(2)}{C}$, which corresponds to
$C$ here, is
\[\stackrel{(2)}{C}=\frac{3h_{1}h_{2}}{2b^{4}}-\frac{m_{1}m_{2}}{b^{2}}, \]
so that (22) gives in this case
\[ p=\frac{m_{1}m_{2}}{(2b)^{2}} - \frac{3h_{1}h_{2}}{8b^{4}}, \]
which includes the expected inverse square law gravitational force.  This
suggests that the argument leading to (22) may be justified.  Formula (22) is
also compatible with the results of other authors [13] [14] [15] in the
purely gravitational case.

\section{Conclusion}
Papapetrou's class of exact solutions includes one for two massless spinning
particles in an axially symmetric configuration.  This contains a
conical singularity between the particles representing a strut balancing the spin-spin
force.  A tentative argument suggests that this force
is given by (23).  The calculations are consistent with those of [10],
in which the interaction was treated by an approximation method.

In [10] it was shown that between massive spinning particles there is, in
general, a {\em torsion singularity} in addition to the conical singularity.
The torsion singularity depends on the masses of the particles as well
as their angular momenta, and does not appear in the solution
of this paper because in it the masses are zero.

The analogy between the gravitational spin-spin force and the force between
two magnetic dipoles has been investigated by Wald [12]. In a subsequent paper
I shall illustrate this analogy by deriving an exact solution of Einstein-
Maxwell theory for two (massless) magnetic dipoles, and comparing it with
the solution in this paper.

\section*{References}
{[1]} Kramer D and Neugebauer G 1980 {\em Phys. Lett. A} {\bf75} 259\\
{[2]} Dietz W and Hoenselaers c 1982 {\em Phys. Rev. Lett.} {\bf48} 778\\
{[3]} Dietz W and Hoenselaers C 1985 {\em Ann. Phys. NY} {\bf 165} 319\\
{[4]} Hoensalaers C 1984 {\em Prog. Theor. Phys.} {\bf72} 761\\
{[5]} Kihara M and Tomimatsu A 1982 {\em Prog. Theor. Phys.} {\bf67} 349\\
{[6]} Dietz W 1984 {\em Solutions of Einstein's Equations} (Berlin: Springer) p85\\
{[7]} Manko V S, Ruiz E and G\'{o}mez 2000 {\em Class. Quantum Grav.} {\bf 17} 3881\\
{[8]} Manko V S and Ruiz E 2001 {\em Class. Quantum Grav.} {\bf 18} L11\\
{[9]} Papapetrou A 1953 {\em Annalen der Physik} {\bf 12} 309\\
{[10]}Bonnor W B 2001 {\em Class. Quantum Grav.} {\bf 18} 1381\\
{[11]}Hiscock W A 1985 {\em Phys. Rev.} D {\bf 31} 3288\\
{[12]}Wald R 1972 {\em Phys. Rev.} D{\bf 6} 406\\
{[13]}Robertson H P and Noonan T W 1968 {\em Relativity and Cosmology}
(Philadelphia: W B Saunders) page 278\\
{[14]}Antoci S, Liebscher D-E and Mihich L 2001 {\em Class. Quantum Grav.} {\bf 18} 3463\\
{[15]}Katz A 1968 {\em J.Math. Phys.} {\bf 9} 983
\end{document}